\documentclass[aps,prl,showpacs,twocolumn]{revtex4}
\usepackage{graphicx}
\input{epsf}

\begin{document}
\title {Macroscopic quantum tunneling in high-$T_c$ superconductors: \\
the role of collective charge oscillations}

\author{M. V. Fistul}

\affiliation {Theoretische Physik III, Ruhr-Universit\"at Bochum,
D-44801 Bochum, Germany}
\date{\today}
\begin{abstract}
We present a theoretical study of an escape rate for switching
from the superconducting state to a resistive one in series arrays
of strongly interacting Josephson junctions. At low temperatures
such a switching is determined by macroscopic quantum tunneling
(MQT) of a spatially extended Josephson phase. An increase of the
crossover temperature from  the thermal to quantum regimes, and a
giant enhancement of MQT escape rate is obtained. Such an effect
is explained by excitation of collective charge oscillations
forming a {\it charge instanton}. Using a model of screened
Coulomb interaction we found that our analysis is in a good accord
with recently published experimental results on an enhancement of
the MQT in single crystals of high-$T_c$ superconductors.
\end{abstract}


\maketitle

Great attention has been devoted to an experimental and
theoretical study of diverse {\it macroscopic quantum phenomena},
e.g. macroscopic quantum tunneling (MQT) and energy level
quantization just to name a few,  in Josephson coupled systems
\cite{Tinkham,Clarke,SQUID-QB1,Kato,Malomed,Sol-QM-Nature,Sol-QM-PRL,HTSC-QM1,HTSC-QM2}.
A lumped Josephson junction is characterized by a single Josephson
phase, and the MQT of a Josephson phase has been found long time
ago in Nb Josephson junction \cite{Tinkham,Clarke}. At low
temperatures the MQT determines the escape rate of the switching
from the superconducting state to a resistive one. The MQT is
considered as a basic effect in the modern field of quantum
information processing \cite{Qcomp}

Next step in a study of macroscopic quantum phenomena is to obtain
the MQT in {\it spatially extended} Josephson systems, i.e. dc
SQUIDs \cite{SQUID-QB1}, quasi-one-dimensional long Josephson
junctions \cite{Kato,Malomed,Sol-QM-Nature,Sol-QM-PRL}, Josephson
parallel and series arrays \cite{HTSC-QM1,HTSC-QM2}. These systems
that can be characterized by coordinate and time dependent
Josephson phase $\varphi(x,t)$, present a particular case of
interacting many particle systems. The presence of inductive
interaction in Josephson parallel arrays and long Josephson
junctions allows one to form diverse macroscopic objects
(solitons), e.g. Josephson vortices (magnetic fluxons) and
vortex-antivortex pairs. Thus, the MQT in spatially extended
Josephson systems can be often reduced to the quantum fluctuation
induced escape of solitons from a pinning potential
\cite{Kato,Malomed,Sol-QM-Nature,Sol-QM-PRL}. Indeed, both the MQT
of a Josephson vortex \cite{Sol-QM-Nature} and the quantum
dissociation of a vortex-antivortex pair \cite{Sol-QM-PRL} have
been observed.

Other example of spatially extended Josephson systems is a dc
biased series array of Josephson junctions. This case presents a
special interest because artificial series arrays of
$Al/Al_20_3/Al$ junctions have been prepared \cite{Haviland}, and
layered high-$T_c$ superconductors can be modelled as a stack of
intrinsic Josephson junctions \cite{Kleiner}. Moreover, modern
fabrication technique allows to prepare single crystals of layered
high-$T_c$ superconductors with an extremely homogeneous
distribution of critical currents of intrinsic Josephson
junctions, and a low level of dissipation
\cite{HTSC-QM1,HTSC-QM2}. Such systems are suitable for an
experimental study of macroscopic quantum phenomena, and indeed,
the MQT of a Josephson phase has been observed in layered
$BiSrCaCuO$ superconductors \cite{HTSC-QM1,HTSC-QM2}.

A rather unexpected result found in the Refs.
\cite{HTSC-QM1,HTSC-QM2} is that the crossover temperature $T^{*}$
from the thermal fluctuation regime to the MQT regime is much
larger in respect to a single Josephson junction having the same
parameters. Moreover, in Ref. \cite{HTSC-QM2} it has been observed
that the MQT escape rate $\Gamma_{MQT}$ for a stack of intrinsic
Josephson junctions is {\it four order of magnitude} larger than
$\Gamma_{MQT}$ for a single Josephson junction. In these
experiments it was also found that the escape rate $\Gamma_T$ in
the thermal fluctuation regime did not differ from the escape rate
of a single Josephson junction.

Since in the model of independent Josephson junctions the
crossover temperature does not depend on number $N$ of junctions,
and the escape rate $\Gamma$ is just proportional to $N$, an
enhancement of the MQT observed in layered high-$T_c$
superconductors stems from an interaction between intrinsic
Josephson junctions. In order to explain such a giant increase of
the MQT escape rate the idea of a long-range coupling between
intrinsic Josephson junctions has been proposed in Ref.
\cite{HTSC-QM2}. Although this mechanism can definitely take
place, we notice that a natural source of interaction in series
junction arrays is the screened Coulomb charge interaction
\cite{Tachicki}. This interaction is especially strong in layered
high-$T_c$ superconductors, where the Debye screening length is of
the order of superconducting layer thickness.

\begin{figure}
\includegraphics[width=2in,angle=-90]{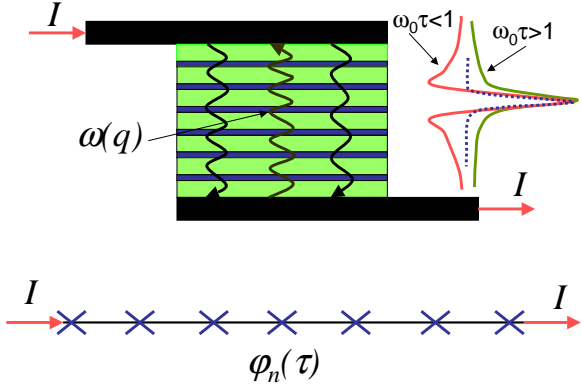}
\caption{Schematic of a dc biased layered high-$T-c$
superconductor and a series array of Josephson junctions.
Collective charge oscillations (wave lines), a strongly localized
instanton (dashed line) and a charge instanton with long tails
(solid lines) are shown.} \label{Schematic}
\end{figure}

In this Letter we will show that the excitation of collective
charge oscillations in a stack of Josephson junctions leads to an
increase of the crossover temperature $T^{*}$ and a giant
enhancement of the MQT escape rate. In order to quantitatively
analyze the escape rate in layered superconductors  a series array
of interacting Josephson junctions biased by dc current $I$ (see
schematic in Fig. 1) is considered.  The system is characterized
by the set of Josephson phases $\varphi_n (\tau)$, where number
$n$ changes from $0$ to $N$. In order to obtain the escape rate
$\Gamma$ we use "instanton technique"
\cite{Coleman,Affleck,Ingold}, and therefore, $\tau$ is the
imaginary time that varies from $0$ to $\hbar/k_b T$ ($T$ is the
temperature). In the imaginary time representation the Lagrangian
of a series array with a charge interaction is written as
$$
L~=~\sum_n \sum_m \frac{1}{2}[\dot \varphi_n (\tau) C_{nm}\dot
\varphi_m (\tau)]+ \sum_n U_n~,
$$
\begin{equation} \label{Lagr}
U_n(\varphi)=\cos \varphi_n (\tau)+i \varphi_n (\tau)~,
\end{equation}
where $ i=I/I_c$ is the normalized external dc current, and $I_c$
is the nominal value of the critical current of a single junction.
The Lagrangian is expressed in units of $E_J$, where $E_J$ is the
Josephson energy of a single junction. The capacitance matrix
$C_{nm}=C_{|n-m|}$ is determined by a charge interaction between
junctions, and $C_{nm}$ becomes the $\delta_{nm}$-function for the
specific case of decoupled junctions, i.e. as the charge
interaction between junctions is small $ C<< C_0$. The $C_0$ is
the capacitance of a single Josephson junction. In the experiments
on the MQT in Josephson coupled systems $E_J>>E_{C_0}~=~e^2/C_0$,
and the MQT occurs as the dc current $I$ is close to $I_c$, and
therefore $(i-1) \ll 1$. In this case  the potential
$U_n(\varphi)$ is written as
\begin{equation} \label{Lagr2}
U_n(\varphi)~=~(1-i) \varphi_n (\tau)-\frac{\varphi_n^3
(\tau)}{6}~.
\end{equation}
The escape rate is determined by the particular solution
$\varphi_n(\tau)$ providing the extremum of action
\begin{equation} \label{Action}
S=-\frac{1}{\hbar}\int_0^{\hbar/k_B T} L(\tau) d\tau ~.
\end{equation}
At high
temperatures such a solution is determined by extremum points of
the potential $U_n$, and it is written as
$$ \varphi_n=2\sqrt{2(1-i)}\delta_{nl}-\sqrt{2(1-i)}~.
$$
Here, $l$ is a junction number where the fluctuation occurs. Since
this solution does not depend on the time $\tau$, we can
immediately conclude that the excitation of charge oscillations
does not change a slope of the dependence of the escape rate
$\ln[\Gamma_{T}(1-i)]$ on the bias current $(1-i)$ in the thermal
fluctuation regime. However, the crossover temperature from the
thermal fluctuation regime to the MQT can be strongly enhanced by
excitation of charge oscillations in a stack of interacting
Josephson junctions. Indeed, using the method elaborated in
\cite{Coleman,Affleck,Ingold} we obtain that at high temperatures
the optimal fluctuation $\varphi_n(\tau)$ around an extremum point
has a form:
\begin{equation} \label{OptflHT}
\varphi_n(\tau)~=~e^{\frac{2\pi i k_B T \tau}{\hbar}} \phi_n~,
\end{equation}
where the eigenfunctions $\phi_n$ are the solution of the nonlocal
and inhomogeneous equation:
\begin{equation} \label{EigenfunctHT}
\sum_m \frac{4\pi^2 k_B^2T^2}{\hbar^2}C_{nm}\phi_m -2\omega_0^2
\delta_{lm}\phi_m ~=~(\lambda-\omega_0^2) \phi_n~.
\end{equation}
Here, $\lambda$ are the eigenvalues of the Eq.
(\ref{EigenfunctHT}), $\omega_0~=~\omega_p[2(1-i)]^{1/4}$ is the
dc bias dependent frequency of oscillations on the bottom of
potential well, $U_n(\varphi)$, and $\omega_p$ is the plasma
frequency of a single Josephson junction. The crossover
temperature $T^*$ is determined by the condition that there is the
eigenvalue $\lambda=0$ \cite{Affleck,Ingold}. Using the Fourier
transform of Eq. (\ref{EigenfunctHT}) the crossover temperature is
obtained as a solution of the following equation:
\begin{equation} \label{Detcrosstemp}
1=\int_0^{2\pi}\frac{dq}{2\pi}\frac{2\omega_0^2}{\omega_0^2+\xi^2[1+2C(q)]}~,
~\xi=\frac{2\pi k_B T^*}{\hbar}~.
\end{equation}
Here, $C(q)$ is the Fourier $\cos$-transform of the function
$C_{nm}$, i.e. $C(q)=\frac{1}{C_0}\sum_{n=1}^{N}C_n\cos(qn)$. We
can analyze  Eq. (\ref{Detcrosstemp}) in different limits. In the
case of decoupled junctions, i.e as $C(q)<<1$ for all $q$, we
obtain that $T_{ind}^*=\hbar \omega_0/(2\pi k_B)$, and the
crossover temperature of a series array $T_{ind}^*$ coincides with
the one for a single Josephson junction. However, for strongly
coupled Josephson junctions $C(q)$ becomes negative for
$q~\simeq~\pi$, and can be of the order of one. In this case, the
crossover temperature has to increase.

This effect can be easily understood if we notice that the
spectrum of charge oscillations in a stack of Josephson junctions
forms a band
\begin{equation} \label{Spectrum}
\omega^2(q)=\frac{\omega_0^2}{1+2C(q)}~.
\end{equation}
The crossover temperature is determined by the {\it maximum} value
of the $\omega(q)$. Therefore, the excitation of high-frequency
charge oscillations leads to an increase of the crossover
temperature. On other hand a well-known effect of an enhancement
of the MQT escape rate under microwave radiation \cite{HTSC-QM2}
is determined in Josephson junction series arrays with a strong
charge interaction by the {\it minimum} value of the $\omega_q$.
Therefore, the value of the plasma frequency found from this
effect is not directly related to the crossover temperature.

Now we turn to the MQT regime, where the extremum point of the
action $S$ is the "tau-dependent" instanton (bounce) solution. In
the absence of a charge interaction the instanton solution is
strongly localized on a particular junction (see schematic in Fig.
1, dashed line), i.e \cite{Ingold}.
\begin{equation} \label{Instantonsol}
\varphi_n(\tau)=f_0(\tau)=\delta_{nl}\frac{3\sqrt{2(1-i)}}{\cosh^2(\omega_0\tau/2)}~~.
\end{equation}
In a generic case of interacting junctions , an instanton dressed
by collective charge oscillations has small tau-dependent spatial
tails (see Fig. 1, solid line). Such a spatial-temporal instanton
solution that can be called a \emph{charge instanton}, is found by
perturbation analysis ($C(q)~\leq~1$) as
\begin{equation} \label{Tailsinstanton}
\varphi_n(t)=\int_0^{2\pi} \frac{dq }{2\pi} \int \int
\frac{d\omega d\tau_1 }{2\pi}
\frac{2C(q)e^{i\omega(\tau-\tau_1)}}{\omega_0^2+\omega^2[1+2C(q)]}
\ddot{f}_0(\tau_1)~~.
\end{equation}
Substituting (\ref{Tailsinstanton}) in the expression for action
$S$ (\ref{Action}) we obtain the escape rate $\Gamma_{MQT}$ (in physical units)
as
\begin{equation} \label{GammaMQT}
\Gamma_{MQT}~=\Gamma_0\exp{\left[-\frac{72 E_J}{15 \hbar
\omega_p}2^{1/4}(1-i)^{5/4}(1-\chi) \right]}~,
\end{equation}
where
\begin{equation} \label{GammaMQT-2}
\chi=60 \pi \int_0^{2\pi} \frac{dq}{2\pi}\int {dx} \frac{C^2(q)
x^6}{\{x^2[1+2C(q)]+1\}\sinh^2(\pi x)}
\end{equation}
The parameter $\chi$ having a positive value, characterizes an
enhancement of MQT due to the presence of the charge interaction
between Josephson junctions, as $C(q)$ is different from zero. The
integral over $x$ can be calculated by taking into account that
the typical values of $x$ are numerically small, i.e $x\leq
(1/\pi)$. Therefore, with a good accuracy we obtain
\begin{equation} \label{CHI-fin}
\chi=\frac{20}{7} \int_0^{2\pi}C^2(q) \frac{dq}{2\pi}
\end{equation}

Thus, remarkably the crossover from the thermal fluctuation regime
to the MQT regime in Josephson series arrays with a charge
interaction is realized as a transition from $0$- to
$2$-dimensional instanton solution.

Presented above a generic analysis can be used for a particular
case of the MQT in a single crystal of high-$T_c$ superconductors
contained intrinsic Josephson junctions. A natural source of a
charge interaction in layered superconductors is the Coulomb
screened charge interaction, and we use a simple model introduced
by M. Tachiki et al. in Ref. \cite{Tachicki}. In this model the
function $C(q)$ has a form:
\begin{equation} \label{Cq}
C(q)=-\frac{\alpha}{2}\frac{1-\cos(q)}{1+\alpha(1-\cos(q))}~,
\end{equation}
where the parameter $\alpha$ characterizes a strength of the
charge interaction. This parameter is determined by the ratio
between the Debye screening length and a superconducting layer
thickness, and $\alpha$ can be larger than one in layered
high-$T_c$ superconductors \cite{Tachicki}. Substituting
(\ref{Cq}) in (\ref{Detcrosstemp}) and (\ref{CHI-fin}), and
calculating the integrals over $q$, we obtain the dependence of
the crossover temperature $T^*$ and the parameter $\chi$ on the
strength of Coulomb interaction $\alpha$:
\begin{equation} \label{tcross-HTSC}
T^*(\alpha)=T^*_{ind}\sqrt{\frac{1+\alpha+\sqrt{4+\alpha^2+8\alpha}}{3}}~
\end{equation}
and
\begin{equation} \label{chi-HTSC}
\chi=\frac{5}{7}\left[1-\frac{1+3\alpha}{(1+2\alpha)^{3/2}}\right]~~.
\end{equation}
These dependencies are shown in Fig. 2.
\begin{figure}
\includegraphics[width=3in,angle=0]{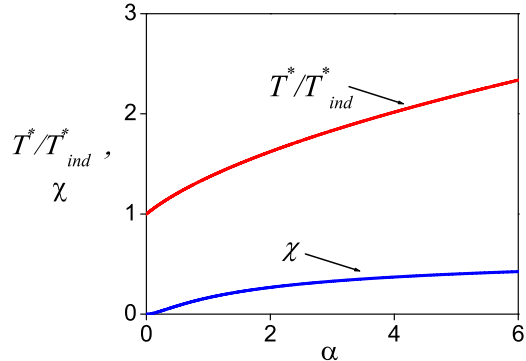}
\caption{The dependencies of the crossover temperature
$T^*(\alpha)$ and the parameter $\chi(\alpha)$ on the strength of
Coulomb interaction $\alpha$. }\label{CrossTdep}
\end{figure}

Moreover, in Fig. 3 we present the dependence of the MQT escape
rate on the external dc bias for various values of $\alpha$. One
can see a giant enhancement of the MQT escape rate as the
parameter $\alpha$ becomes larger than one. This enhancement
results from a decrease of the slope of the bias current
dependence escape rate $\ln[\Gamma_{T}(1-i)]$ on  $(1-i)$.
Comparing our theoretical predictions with the experimental curves
published in Ref. \cite{HTSC-QM2} (see Fig. 5 in Ref.
\cite{HTSC-QM2}) we find a good agreement as $\alpha~\simeq~4$. By
making use of the same value of $\alpha$ we obtain that the
crossover temperature increases two times in respect to a single
Josephson junction with the same parameters.

Notice here that for a finite system as a number of intrinsic
Josephson junctions $N$ is not very large, there is a discrete
number of collective oscillation modes that can be excited in a
system. In this case the integral over $q$ in Eq. (\ref{CHI-fin})
has to be changed to the sum over discrete modes, i. e. $\int
dq/(2\pi) \rightarrow (1/N)\sum_n$. Therefore, the slope of the
bias current dependence escape rate $\ln[\Gamma_{T}(1-i)]$
increases for a series array with less number of Josephson
junctions. It gives a natural explanation of a moderate
enhancement of the MQT escape rate as a number of intrinsic
Josephson junction $N$ was decreased \cite{HTSC-QM2}.

\begin{figure}
\includegraphics[width=3in,angle=0]{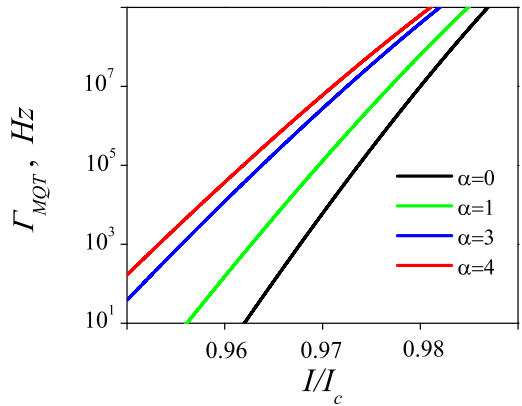}
\caption{The dependence of the MQT escape rate $\Gamma_{MQT}$ on
the dc bias current $I$ for various values of $\alpha=0,~1,~3,~4$.
Legend corresponds to the curves arranged from right to left. In
order to fit our theoretical predictions to the experimental
results of Ref. \cite{HTSC-QM2} the value of $E_J/(\hbar
\omega_p)~\simeq~264$ and preexponent in the expression
(\ref{GammaMQT}) $\Gamma_0~=~1~GHz$ were used. }\label{MQTratedep}
\end{figure}

In conclusion we have shown that the MQT escape rate can be
drastically increased in strongly coupled Josephson junctions
series arrays. The MQT in such arrays is described through the
quantum fluctuation induced excitation of a charge instanton
characterized by long spatial tails. It is in marked contrast to
the well-known cases of a single Josephson junction or weakly
coupled Josephson junctions series arrays where a strongly
localized instanton solution forms. Appearance of the charge
instanton is explained by excitation of collective charge
oscillations, and this mechanism has to be especially effective in
layered high-$T_c$ superconductors where the screening of Coulomb
interaction is rather weak. The excitation of high-frequency
charge oscillations also leads to a significant increase of the
crossover temperature from the thermal fluctuation to the quantum
regime. Such an enhancement of the MQT in Josephson series arrays
with a strong charge interaction can be very promising for a field
of quantum information processing \cite{Qcomp}.

I would like to thank Yu. Koval, P.M\"uller, A. Volkov and A. V.
Ustinov for useful discussions and attraction of my attention to
the MQT problem in high-$T_c$ superconductors. I acknowledge the
financial support by SFB 491.


\begin{thebibliography}{99}

\bibitem{Tinkham} M. Tinkham, {\it Introduction to Superconductivity}, 2nd edition,
McGraw-Hill, New York (1996)

\bibitem{Clarke} R. F. Voss and R. A. Webb, Phys. Rev. Lett.
\textbf{47}, 265 (981); J. M. Martinis and J. Clarke, Phys. Rev.
Lett. \textbf{55}, 1908 (1985).

\bibitem{SQUID-QB1} F. Balestro, J. Claudon, J. P. Pekola, and O.
Buisson,
Phys. Rev. Lett. \textbf{91}, 158301 (2003); S.-X. Li, Y. Yu, Y.
Zhang, W. Qiu, S. Han, and Z. Wang, Phys. Rev. Lett. 89, 098301
(2002)

\bibitem{Kato} T. Kato and M. Imada, J. Phys. Soc. Jpn.
\textbf{65}, 2963 (1996); T. Kato, J. Phys. Soc. Jpn., \textbf{69}
2735 (2000).

\bibitem{Malomed} A. Schnirman, E. Ben-Jacob, and B. A. Malomed,
Phys. Rev. B. \textbf{56}, 14677 (1997).

\bibitem{Sol-QM-Nature} A. Wallraff, A. Lukashenko, J. Lisenfeld,
A. Kemp, Yu. Koval, M. V. Fistul, and A. V. Ustinov, Nature
\textbf{425}, 155 (2003).

\bibitem{Sol-QM-PRL} M. V. Fistul, A. Wallraff, Yu. Koval, A. Lukashenko,
B. A. Malomed, and A. V. Ustinov, Phys. Rev. Lett. \textbf{91},
257004 (2003).

\bibitem{HTSC-QM1} K. Inomata, S. Sato, Koji Nakajima,
 A. Tanaka, Y. Takano, H. B. Wang, M. Nagao, H. Hatano, and S. Kawabata,
 Phys. Rev. Lett. \textbf{95}, 107005
(2005).

\bibitem{HTSC-QM2} X. Y. Jin, J. Lisenfeld, Y. Koval, A.
Lukashenko, A. V. Ustinov, and P. M\"uller, Phys. Rev. Lett.
\textbf{96}, 177003 (2006).

\bibitem{Qcomp} J. M. Martinis, S. Nam, J. Aumentado, and C. Urbina,
Phys. Rev. Lett. \textbf{89}, 117901 (2002); R. McDermott et al.,
Science \textbf{307}, 1299 (2005)

\bibitem{Haviland} E. Chow, P. Delsing, and D. B. Haviland,
Phys. Rev. Lett. \textbf{81}, 204 (1998).

\bibitem{Kleiner} R. Kleiner, F. Steinmeyer, G. Kunkel, and P. M\"uller, Phys. Rev. Lett.
\textbf{68}, 2394 (1992); R. Kleiner and P. M\"uller, Phys. Rev. B
\textbf{49}, 1327 (1994).

\bibitem{Tachicki} T. Koyama and M. Tachiki, Phys. Rev. B,
\textbf{54}, 16183 (1996); A. Gurevich and M. Tachiki, Phys. Rev.
Lett., \textbf{83}, 183 (1999).

\bibitem{Coleman} S. Coleman, Phys. Rev. D \textbf{15}, 2929
(1977);C. G. Callan, Jr. and S. Coleman, Phys. Rev. D \textbf{16},
1762 (1977).

\bibitem{Affleck} I. Affleck, Phys. Rev. Lett. \textbf{46}, 388
(1981)

\bibitem{Ingold} G-L. Ingold in \emph{Quantum Transport and Dissipation}, Th.
Dittrich, G-L. Ingold, G. Sch\"on, P. H\"anggi, B. Kramer, and W.
Zwerger, , Wiley-VCH (1998).


\end{thebibliography}
\end{document}